\documentclass[conference]{IEEEtran}
\IEEEoverridecommandlockouts
\usepackage{cite}
\usepackage{amsmath,amssymb,amsfonts}

\counterwithin*{theorem}{section}

\usepackage{hyperref}
\hypersetup{
	colorlinks=true,
	linkcolor=blue,
	filecolor=magenta,
	urlcolor=cyan,
}

{
	\newtheorem{assumption}{Assumption}
	
}
\newtheorem{remark}{Remark}

\usepackage{algorithmic}
\usepackage{float}
\usepackage{graphicx}
\usepackage{textcomp}
\usepackage{xcolor}
\def\BibTeX{{\rm B\kern-.05em{\sc i\kern-.025em b}\kern-.08em
    T\kern-.1667em\lower.7ex\hbox{E}\kern-.125emX}}
\makeatletter
\let\old@ps@headings\ps@headings
\let\old@ps@IEEEtitlepagestyle\ps@IEEEtitlepagestyle
\def\confheader#1{%
	\def\ps@IEEEtitlepagestyle{
		\old@ps@IEEEtitlepagestyle
		\def\@oddhead{\strut\hfill#1\hfill\strut}
		\def\@evenhead{\strut\hfill#1\hfill\strut}
	}
	\ps@headings
}
\makeatother
\confheader{
	\small{Proceedings of the 10th RSI International Conference on Robotics and Mechatronics (ICRoM 2022), Nov. 17-19, 2022, Tehran, Iran}
}
\usepackage[pscoord]{eso-pic}
\newcommand{\placetextbox}[3]{
	\setbox0=\hbox{#3}
	\AddToShipoutPictureFG*{ \put(\LenToUnit{#1\paperwidth},\LenToUnit{#2\paperheight}){\vtop{{\null}\makebox[0pt][c]{#3}}}
	}
}
\placetextbox{.23}{0.055}{\textbf{\small{978-1-6654-2094-5/21/\$31.00~\copyright 2021 IEEE}}}
\begin{document}

\title{Reinforcement learning-based optimised control for tracking of nonlinear systems with adversarial attacks\\
}

\author{\IEEEauthorblockN{ Farshad Rahimi }
\IEEEauthorblockA{\textit{Department of Electrical Engineering } \\
\textit{  Sahand University of Technology, Tabriz}\\
PO. BOX 51335/1996, Iran. \\
email address: fa\_rahimi@sut.ac.ir}
\and
\IEEEauthorblockN{Sepideh Ziaei}
\IEEEauthorblockA{\textit{Department of Electrical Engineering } \\
\textit{Sahand University of Technology}\\
Tabriz, Iran.\\
email address: s\_ziyaei400@sut.ac.ir}

}

\maketitle

\begin{abstract}
This paper introduces a reinforcement learning-based tracking control approach for a class of nonlinear systems using neural networks. In this approach, adversarial attacks were considered both in the actuator and on the outputs. This approach incorporates a simultaneous tracking and optimization process. It is necessary to be able to solve the Hamilton-Jacobi-Bellman equation (HJB) in order to obtain optimal control input, but this is difficult due to the strong nonlinearity terms in the equation. In order to find the solution to the HJB equation, we used a reinforcement learning approach.  In this online adaptive learning approach, three neural networks are simultaneously adapted: the critic neural network, the actor neural network, and the adversary neural network. Ultimately, simulation results are presented to demonstrate the effectiveness of the introduced method on a  manipulator.
\end{abstract}


\section{Introduction}
In the past few years, the trajectory tracking problem has been increasingly investigated for nonlinear systems.  As one of the kinds of nonlinear systems to model various engineering systems, the second-order mathematical model has been attracting more and more attention due to its wide applications, such as the robotic platform, unmanned aerial vehicle \cite{ff1,ff2}. There have been a number of publications on the tracking control problem topic, including \cite{h1,h2,h3}.
For example, as a result of transforming the non-affine uncertain nonlinear systems with input saturation into equivalent affine models, two methods are proposed in \cite{ff3}. In one approach, linearisation techniques are used to obtain relevant system inputs; in the other approach, the Filippov model and pulse width modulation are used to create an average dynamics model.  Taking the desired trajectory and velocity as a virtual leader is one of the main advantages of the tracking control problem. Also, in \cite{ff4}, researchers employ artificial potentials to approach the obstacle avoidance problem of nonlinear multi-agent formation control.  The solution to tracking and controlling nonlinear systems has recently been investigated using optimal control based on reinforcement learning.
\\
Optimal control involves finding a control policy that minimizes a predefined performance index of the dynamic system, balancing control task and resources. There has always been a lot of research going on in the area of optimal control. There have been numerous applications of optimal theory, including industrial processes, aerospace, robotics, and vehicles \cite{hr2}. As a powerful tool for solving optimization problems, reinforcement learning has attracted a great deal of interest within the control community \cite{ff5,hr3}.
Reinforcement learning refers to evaluating feedback from the environment to produce appropriate control behaviors, which performs actor–critic analysis.  Neural networks are efficient and popular tools in nonlinear control because of their universal approximation, learning, and adaptation capabilities \cite{hr4,hr5}. The reinforcement learning-based nonlinear optimal control is further investigated by using neural networks to estimate solutions to the Hamilton–Jacobi–Bellman (HJB) equation, and many highlighted results have been proposed in the past few decades \cite{hr6}.
The core structure applied to the implementation of reinforcement learning is the actor-critic architecture \cite{ff6}.  For a class of stochastic systems with asymmetric actuator dead zones, \cite{hr7} proposes an adaptive nonlinear tracking control scheme utilizing neural networks backstepping. The authors of \cite{hr10} study proposed an adaptive neural network-based reinforcement learning approach for optimizing tracking control for nonlinear systems.  Also, the finite-horizon optimal control problem of continuous-time nonlinear systems with unknown draft system dynamics is investigated in \cite{hr11}.
\\
Another issue that has recently been discussed in control problems as it pertains to reading data from sensors is cyber-threats. We should consider these threats when designing controllers for our systems. As a result of the vulnerability of the communication networks and sensors and the ability to hack and modify them, the performance of the systems can be compromised \cite{L6}. Through cyberattacks, the attacker attempts to compromise the integrity of the system. By using deception information, incorrect sensor measurements, or false data injection, these attacks are often undertaken. Therefore, in the process of designing a control system or analyzing stability, cyber-attacks need to be taken into account. A number of effective strategies have been proposed in the literature to defend against these types of attacks \cite{L7,L10}. A system with attack/fault tolerance was proposed in \cite{L10}. False data injection attacks are studied in \cite{L13} for event-triggered security consensus for a class of multiagent systems. The authors present a mechanism for adaptive event triggering for a class of linear multi-agent systems that considers cyberattacks \cite{L12}.
\\
As a result of the difficulty in controlling and analyzing the system's convergence, existing optimal control methods in the presence of adversarial attacks rarely address tracking control problems . The purpose of this paper is to introduce a reinforcement learning-based optimal control scheme for a class of nonlinear systems under cyberattacks on actuators and output measurements in light of the aforementioned literature as well as \cite{hr10,hr11}.
The main contributions of this paper can be highlighted as follows:\\
(1) In the presence of adversarial attacks, an optimized control approach is provided to solve the nonlinear tracking problem efficiently. The tracking error term from the optimal performance index function is segmented in order to guarantee both trajectory tracking and performance optimization. The actor-critic reinforcement learning method is used to obtain an admissible control input. \\
(2) It is possible to apply the proposed method to optimize nonlinear systems with control gain functions in a general way, and so it can be applied to more real-world systems.
\\
This paper is organized as follows.  A tracking control problem is formulated for nonlinear systems that takes into account cyberattacks during sensor reading in Section \ref{sec2}. In section \ref{sec3}, simulation results for a robotic arm are provided. Ultimately, the conclusion is given in Section \ref{sec4}.\\

\section{PROBLEM FORMULATION AND PRELIMINARIES }
\label{sec2}
\subsection{Plant description}
Consider the following nonlinear system:
\begin{align}
		 \dot{x}\left( t \right)=f\left( x \right)+g\left( x \right)u(t)+\beta \left( x \right)v(t),
		 \label{eq1}
\end{align}
where $x\left( t \right)\in {{R}^{n}}$ stands for the states of the system , $u(t)\in {{R}^{n}}$ and $v(t)\in {{R}^{n}}$ represent the control input and the adversarial attacks, respectively. $f\left( x \right)\in {{R}^{n}}$ with $f\left( 0 \right)={{0}_{n}}$ is the dynamic function, $g\left( x \right)\in {{R}^{n\times n}}$ is the control gain function, $\beta \left( x \right)\in {{R}^{n\times n}}$ is the adversarial gain function. The term $f\left( x \right)+g\left( x \right)u(t)$ supposes to Lipschitz continuous. That will guarantee (\ref{eq1}) in order to the existence of a unique solution for the bounded initial value. There exists a $u(t)$ that makes the system has the  asymptotic stability, i.e., the system (\ref{eq1}) is stabilizable.  As a first step, the following assumption must be made for establishing the presented approach.
\begin{assumption}
The system dynamic function $f\left( x \right)~$ and control gain function $g\left( x \right)$ are known and bounded, and the continuous matrix function $g\left( x \right)$ is nonsingular, thus making $g\left( x \right)$ a matrix that can be inverted.
\label{ass1}
\end{assumption}
\subsection{The tracking control formulation}
Let ${{\eta }_{d}}(t)\in {{R}^{n}}~$ stand for the desired tracking trajectory, and then the tracking error is defined as  $\eta (t)=x(t)-{{\eta }_{d}}(t)$. The variables $~{{\eta}_{d}}\left( t \right)~$ and ${{\dot{\eta}}_{d}}\left( t \right)~$ suppose  to be bounded. Based on the system (\ref{eq1}), we obtain the following equation:
\begin{align}
\dot{\eta}(t)=f(x)+g(x)u(t)+\beta(x)v(t)-{{\dot{\eta}}_{d}}(t),
\label{eq2}
\end{align}
in order to determine the performance index based on (\ref{eq2}), define it as follows:
\begin{align}
  & J(\eta)=\int_{t}^{\infty }h( \eta(t),u(t),v(t))dp ,
 \label{eq3}
\end{align}
where $h(\eta(t),u(t),v(t))={{\eta}^{T}}(t){R}(x)\eta(t)+{{u}^{T}}(t)u(t)-{{\gamma }^{2}}{{v}^{T}}(t)v(t)~$ is immediate or local cost function, and $R(x)=g(x)g^{T}{{(x)}}\in {{R}^{n\times n}}$, in accordance with Assumption \ref{ass1}, is positive definite matrix. $\gamma$ is a positive scalar.\\
Our optimal tracking control problem involves finding a control policy $u(t)$ that minimizes the performance index (\ref{eq3}). In other words,  the optimal control input should be admissible.
Control protocols are admissible on $\Omega$ if they are continuous, and $u\left( 0 \right)=0$, stabilize (\ref{eq1}), and make $J\left( \eta \right)~$ finite on $ \Omega $ denoted by $u(t) \in \Psi (\Omega) $.
\subsection{The control objective}
In this paper, we aim to find the optimized control input $u(t)\in \text{ }\!\!\Psi\!\!\text{ }\left( \text{ }\!\!\Omega\!\!\text{ } \right)$ for (\ref{eq1}) such that all error signals would be semi-globally uniformly ultimately bounded, and the output of the system can follow the predefined trajectory ${{\eta}_{d}}\left( t \right)$ accurately.\\
Associated with (\ref{eq2}) and (\ref{eq3}), Hamiltonian function is generated as
\begin{align}
   H\left( \eta(t),u(t),v(t),J(\eta) \right)=&{\eta}^{T}{{(t)}}R(x)\eta(t)+{{u}^{T}}(t)u(t)\cr &-{{\gamma }^{2}}{{v}^{T}}(t)v(t)
  +\frac{\partial J\left( \eta \right)}{\partial \eta}\Big( f\left( x \right)\cr & +g\left( x \right)u(t)-{{{\dot{\eta}}}_{d}}\left( t \right) \cr &+{\beta }(x)v~(t)\Big),
  \label{eq4}
\end{align}
where ${{J}_{\eta}}(\eta)=\partial J(\eta)/\partial \eta~$ is the gradient of $~J(\eta)$ with respect to $\eta(t)$.\\
Let ${{u}^{*}}(t)~$ represents the optimal control and ${{v}^{*}}(t)$ represent the worst case of adversarial attacks:
\begin{align}
  & {{u}^{*}}(t)=\underset{u(t)\in \text{ }\!\!\Omega\!\!\text{ }}{\mathop{\min }}\,\underset{v(t)}{\mathop{\max }}\,(\int_{t}^{\infty }{h(\eta(t),u(t),v(t))dp } \cr
 & \,\,\,\,\,\,\,\,\,\,\,\,=\int_{t}^{\infty }{h(\eta(t),{{u}^{*}}(t),{{v}^{*}}(t))dp },
 \label{eq55}
\end{align}
The optimal performance index (\ref{eq55}) provides the following HJB equation using the optimal performance index:
\begin{align}
   H(\eta(t),{{u}^{*}}(t),{{v}^{*}}(t),&{{J}^{*}}(\eta))=\eta^{T}{{(t)}}R(x)\eta(t)+{{u}^{*}}^{T}{{(t)}}{{u}^{*}}(t) \cr
 & -{{\gamma }^{2}}{{v}^{*}}^{T}{{(t)}}{{v}^{*}}(t)+{J_{\eta}^{*}}^{T}{{(t)}}\Big(f(x) \cr
 &+g(x){{u}^{*}}(t)+{\beta}(x){{v}^{*}}(t)-{{\dot{\eta}}}_{d}(t)\Big),
 \label{eq6}
\end{align}
where ${{J}_{\eta}}^{*}(\eta)=\partial {{J}^{*}}(\eta)/\partial \eta$.
As a result of solving the differential equation, the optimal control ${{u}^{*}}(t)$ and ${{v}^{*}}(t)$ can be obtained as
\begin{align}
&\frac{\partial H( \eta(t),{{u}^{*}}(t),{{v}^{*}}(t),{{J}_{\eta}}^{*}(\eta))}{\partial {{u}^{*}}(t)}=0, \cr & \Rightarrow ~~~{{u}^{*}}(t)=-\frac{1}{2}{{g}^{T}}(x)J_{\eta}^{*}( \eta),
\label{eq7}
\end{align}
\begin{align}
&\frac{\partial H( \eta(t),{{u}^{*}}(t),{{v}^{*}}(t),{{J}_{\eta}}^{*}(\eta))}{\partial {{v}^{*}}(t)}=0, \cr & ~\Rightarrow ~~~{{v}^{*}}(t)=\frac{1}{2}{{\beta}^{T}}( x)J_{\eta}^{*}( \eta).
\label{eq8}
\end{align}
Substituting (\ref{eq7})-(\ref{eq8}) into (\ref{eq6}), one can have
\begin{align}
	& H( \eta(t),{{u}^{*}}(t),{{v}^{*}}(t),J^{*}{(\eta)})=\eta^{T}{{(t)}}R(x)\eta(t)\cr &+{{\Big( -\frac{1}{2}{g}^{T}{{( x)}}J_{\eta}^{*}(\eta)\Big)}}^{T}\Big( -\frac{1}{2}g^{T}{{( x)}}J_{\eta}^{*}( \eta)\Big) \cr
	& -{{\gamma }^{2}}{{\Big( \frac{1}{2}\beta^{T}{{( x)}}J_{\eta}^{*}( \eta)\Big)}}^{T}\left( \frac{1}{2}\beta^{T}{{( x )}}J_{\eta}^{*}( \eta) \right)+{J_{\eta}^{*}}^{T}{{( \eta)}}f( x ) \cr
	& -{J_{\eta}^{*}}^{T}{{( \eta)}}{{{\dot{\eta}}}_{d}}(t)+{J_{\eta}^{*}}^{T}{{( \eta)}}g\left( x \right)\left( -\frac{1}{2}g^{T}{{( x )}}J_{\eta}^{*}\left( \eta \right) \right) \cr
	& +{J_{\eta}^{*}}^{T}{{(\eta)}}\beta( x){{(\frac{1}{2}{\beta}^T \left( x \right)}}J_{\eta}^{*}{{\left( \eta \right)}}).
	\label{eq9}
\end{align}
The equation (\ref{eq9} can be rewritten as follows:
\begin{align}
  & H( \eta (t),{{u}^{*}}(t),{{v}^{*}}(t),{{J}^{*}}(\eta ))={{\eta }^{T}}(t)R(x)\eta(t) \cr
 & \,\,\,\,\,\,\,\,\,\,\,\,\,\,\,\,\,\,\,\,\,\,\,\,\,\,\,\,\,\,\,\,\,\,+{J_{\eta}^{*}}^{T}{{(\eta )}}f(x)-{J_{\eta }^{*}}^{T}{{(\eta)}}{{{\dot{\eta }}}_{d}}(t) \cr
 & \,\,\,\,\,\,\,\,\,\,\,\,\,\,\,\,\,\,\,\,\,\,\,\,\,\,\,\,\,\,\,\,\,\,-\frac{1}{4}{J_{\eta }^{*}}^{T}{{(\eta )}}R( x )J_{\eta }^{*}(\eta )=0.
 \label{eq10}
\end{align}
It is possible to calculate the gradient term  ${{J}_{\eta}}^{*}(\eta)$ by solving equation (\ref{eq10}); then, the optimal solution can be obtained by inserting the solution into equation (\ref{eq7})-(\ref{eq8}).  Because of the strong nonlinearity of the equation, analytical methods are difficult or even impossible to solve. To overcome the difficulty, it may be possible to use a neural network-based actor-critic RL algorithm.
\subsection{Optimized control design based on RL}
\begin{figure}[h]
	\centering
	\includegraphics[width=0.98\linewidth,height=0.2\textheight]{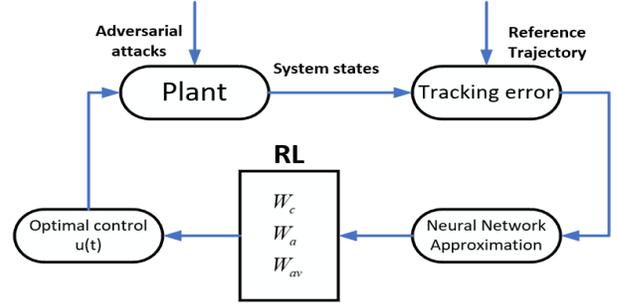}
	\caption{Framework of the proposed method.}
	\label{fig1}
\end{figure}

The purpose of this subsection is to solve the equation (\ref{eq10}) using RL algorithm.
The structure of the proposed approach can be see in Fig. \ref{fig1}.
 It is essential to decompose the optimized performance function into two parts in order to construct the optimal tracking control as follows:
\begin{align}
{{J}^{*}}( \eta )&=\delta  {{\left| \left| \eta\left( t \right) \right| \right|}^{2}}-\beta {{\left| \left| \eta\left( t \right) \right| \right|}^{2}}+{{J}^{*}}( \eta)\cr &=\beta {{\left| \left| \eta\left( t \right) \right| \right|}^{2}}+{{J}^{0}}( \eta).
\end{align}
In which $\delta >0$ is a design constant, ${{J}^{0}}( \eta )={{J}^{*}}( \eta)-\delta {{\left\| \eta(t) \right\|}^{2}}$ .
Due to neural networks’ ability to approximate continuous functions on desired accuracy, specified the compact set ${{\Omega }_{\eta}}$, for $\forall \eta\in {{\Omega }_{\eta}}$ , the continuous function can be approximated as follows:
\begin{align}
	{{J}^{0}}(\eta)={{W}^{*}}^{T}S( \eta)+\xi ( \eta ),
	\label{eq12}
\end{align}
in which ${{W}^{*}}\in {{R}^{m}}$ is the optimal weight matrix, $S\left( \eta \right)\in {{R}^{m}}$is the basis function vector, and  is the neuron number. Based on the neural network approximation (\ref{eq12}), we can rewrite $J_{\eta}^{*}(\eta)$ and ${{u}^{*}}(t)$ optimal control as follows:
\begin{align}
	J_{\eta}^{*}=2\delta \eta(t)+\frac{{{\partial }^{T}}S( \eta)}{\partial \eta}{W_{c}^{*}}(t)+\frac{\partial \xi( \eta)}{\partial \eta},
\label{eq122}
\end{align}
\begin{align}
  & {{u}^{*}}(t)=-\delta {{g}^{T}}(x)\eta(t)-\frac{1}{2}{{g}^{T}}(x)\frac{{{\partial }^{T}}S( \eta)}{\partial \eta }W_{a}^{*}(t) \cr
 & \,\,\,\,\,\,\,\,\,\,\,\,\,\,\,\,\,\,-\frac{1}{2}{{g}^{T}}{{(x)}}\frac{\partial \xi (\eta)}{\partial \eta },
\label{eq14}
\end{align}
\begin{align}
  & {{v}^{*}}(t)=\delta {{\beta }^{T}}(x)\eta(t)+\frac{1}{2}{{\beta }^{T}}(x)\frac{{{\partial }^{T}}S(\eta)}{\partial \eta }W_{av}^{*}(t) \cr
 & \,\,\,\,\,\,\,\,\,\,\,\,\,\,\,\,-\frac{1}{2}{{\beta }^{T}}(x)\frac{\partial \xi(\eta)}{\partial \eta} ,
\label{eq15}
\end{align}
where $\frac{{{\partial }^{T}}S\left( \eta \right)}{\partial \eta}\in {{R}^{n\times m}}$and $\frac{\partial \xi \left( \eta \right)}{\partial \eta}\in {{R}^{n}}$ is the gradient with respect to $\eta$. $~{{{W}}_{c}^{*}}\left( t \right)\in {{R}^{m}}$,  $~{{{W}}_{av}^{*}}\left( t \right)\in {{R}^{m}}$, and ${{{W}}_{a}^{*}}\left( t \right)\in {{R}^{m}}$ would be defined later. \\
Now, by substituting (\ref{eq14})-(\ref{eq15}) into (\ref{eq4}), we can have:
\begin{align}
  & H(\eta (t),u(t),v(t),J(\eta ))=\eta^{T}{{(t)}}R(x)\eta(t)
  +\Big(-\delta {{g}^{T}}(x)\eta(t)\cr &-\frac{1}{2}{{g}^{T}}(x)\frac{{{\partial }^{T}}S(\eta)}{\partial \eta }W_{a}^{*}(t)-\frac{1}{2}{{g}^{T}}(x)\frac{\partial \xi (\eta)}{\partial \eta }\Big)^{T}\Big(-\delta {{g}^{T}}(x)\eta ( t )\cr &-\frac{1}{2}g^{T}{{( x)}}\frac{{{\partial }^{T}}S ( \eta )}{\partial \eta }W_{a}^{*}(t)-\frac{1}{2}g^{T}{{( x )}}\frac{\partial \xi ( \eta  )}{\partial \eta }\Big)
  -{{\gamma }^{2}}\Big(\delta {{\beta }^{T}}(x)\cr& \times \eta (t)+\frac{1}{2}{{\beta }^{T}}( x )\frac{{{\partial }^{T}}S( \eta )}{\partial \eta }W_{av}^{*}(t)-\frac{1}{2}{{\beta }^{T}}(x)\frac{\partial \xi ( \eta )}{\partial \eta }\Big)^{T} \cr
 & \times \Big(\delta {{\beta }^{T}}( x)\eta (t)+\frac{1}{2}{{\beta }^{T}}{{( x)}}\frac{{{\partial }^{T}}S( \eta )}{\partial \eta }W_{av}^{*}(t)\cr& -\frac{1}{2}{{\beta }^{T}}( x)\frac{\partial \xi ( \eta )}{\partial \eta }\Big)+{\Big(2\delta \eta( t)+\frac{{{\partial }^{T}}S(\eta  )}{\partial \eta }W_{c}^{*}(t)+\frac{\partial \xi (\eta)}{\partial \eta}\Big)^{T}}\cr & \times \Big(f(x)+g(x)u(t)-{{{\dot{\eta }}}_{d}}(t)
 +\beta (x)v~(t)\Big)=0.
\end{align}
Due to the unknown matrix of the neural network weights $~{{W_{a}}^{*}}(t), ~{{W_{av}}^{*}}(t), ~{{W_{c}}^{*}}(t)$, the optimal controller (\ref{eq14}) cannot be calculated. An RL algorithm is employed in actor-critic architecture to achieve tracking control. In light of (\ref{eq122}) and (\ref{eq15}), we can design the actor and critic as follows:
\begin{align}
	\hat{J}_{\eta}^{*}\left( \eta \right)=2\delta \eta\left( t \right)+\frac{{{\partial }^{T}}S\left( \eta \right)}{\partial \eta}\hat{W}_{c}^{T}(t),
	\label{eq18}
\end{align}
 \begin{align}
 	{{\hat{u}}^{*}}(t)=-\delta g^{T}{{(x)}}\eta( t)-\frac{1}{2}g^{T}{{( x)}}\frac{{{\partial }^{T}}S( \eta)}{\partial \eta}\hat{W}_{a}^{T}(t),
 	\label{eq19}
 \end{align}
 \begin{align}
 	{{\hat{v}}^{*}}(t)=-\delta \beta^{T}{{(x)}}\eta(t)-\frac{1}{2}\beta^{T} {{(x)}}\frac{{{\partial }^{T}}S(\eta)}{\partial \eta}\hat{W}_{av}^{T}(t),
 	\label{eq20}
 \end{align}
 where ${{\hat{J}}^{*}}(\eta)$ denotes estimation of $~{{J}^{*}}(\eta)$ , and ${{\hat{W}}_{c}}(t)\in {{R}^{m}}$,  and ${{\hat{W}}_{a}}(t)\in {{R}^{m}}$ are critic and actor neural network weight vectors.
The approximated HJB equation is derived by inserting (\ref{eq18})-(\ref{eq20}) into (\ref{eq4}) as follows:
\begin{align}
& H(\eta(t),{{{\hat{u}}}^{*}}(t),{{{\hat{v}}}^{*}}(t),\frac{\partial {{{\hat{J}}}^{*}(\eta)}}{\partial \eta})=\eta^{T}{{(t)}}R(x)\eta(t)\cr&+\delta g^{T}{{(x)}}\eta(t) +\frac{1}{2}g{{\left( x \right)}^{T}}\frac{{{\partial }^{T}}S( \eta)}{\partial \eta}\hat{W}_{a}^{T}{{(t)}} -{{\gamma }^{2}}\delta \beta^{T}{{(x)}}\eta(t) \cr
&-\frac{1}{2}\beta^{T}{{(x)}}\frac{{{\partial }^{T}}S(\eta)}{\partial \eta}\hat{W}_{av}^{T}{{(t)}} +{{\left( 2\beta \eta( t)+\frac{{{\partial }^{T}}S(\eta)}{\partial \eta}{{{\hat{W}}}_{c}}(t) \right)}^{T}} \cr
& \,\,\,\times \Big(f(x)+g(x)\delta g^{T}{{(x)}}\eta(t) -\frac{1}{2}g( x)g^{T}{{(x)}}\frac{{{\partial }^{T}}S(\eta)}{\partial z}{{{\hat{W}}}_{a}}(t) \cr
& -\delta \beta(x)\beta^{T}{{(x)}}\eta(t)-\frac{1}{2}\beta(x)\beta^{T}{{( x)}}  \frac{{{\partial }^{T}}S(\eta)}{\partial \eta}{{{\hat{W}}}_{av}}(t)\cr &-{{{\dot{\eta}}}_{d}}(t)\Big).
\end{align}
Now we can describe Bellman residual error $\phi(t)$ as
\begin{align}
  & \phi \left( t \right)=H\left( \eta (t),u(t),v(t),\frac{\partial {{{\hat{J}}}^{*}}(\eta )}{\partial \eta } \right) \cr
 & \,\,\,\,\,\,\,\,\,\,\,\,\,\,-H\left( \eta (t),{{u}^{*}}(t),{{v}^{*}}(t),\frac{\partial {{J}^{*}}(\eta )}{\partial \eta } \right) \cr
 & \,\,\,\,\,\,\,\,\,\,=H\left( \eta (t),u(t),v(t),\frac{\partial {{{\hat{J}}}^{*}}(\eta )}{\partial \eta } \right).
\end{align}
 Gradient descent method is used to derive the critic updating law for the goal of minimizing Bellman residual error. Defining a positive function as $\text{ }\!\!~\!\!\text{  }\!\!\Phi\!\!\text{ }\left( \text{t} \right)={{\phi }^{2}}\left( t \right)$, critic updating law is produced as follows:
\begin{align}
	& {{{\hat{W}}}_{c}}(t)=-{{\alpha }_{c}}\frac{\partial \Phi (t)}{\partial {{{\hat{W}}}_{c}}(t)}=-{{\alpha }_{c}}\frac{\partial \Phi (t)}{\partial \varphi (t)}\frac{\partial \varphi (t)}{\partial {{{\hat{W}}}_{c}}(t)} \cr
	& \,\,\,\,\,\,\,\,\,\,\,=-{{\alpha }_{c}}2\varphi (t)\frac{\partial \varphi (t)}{\partial {{{\hat{W}}}_{c}}(t)}=-{{\alpha }_{c}}2\varphi (t)\frac{{{\partial }^{T}}J(\eta)}{\partial \eta(t)} \cr
	& \,\,\,\,\,\,\,\,\,\,\,\,\,\,\times \Big(f(x)-\delta g(x)g^{T}{{(x)}}\eta(t)-\frac{1}{2}g(x)g^{T}{{(x)}} \cr
	& \,\,\,\,\,\,\,\,\,\,\,\,\,\,\times \frac{{{\partial }^{T}}J(\eta)}{\partial \eta}{{{\hat{W}}}_{a}}(t)-\delta \beta (x)\beta^{T}{{(x)}}\eta(t) \cr
	& \,\,\,\,\,\,\,\,\,\,\,\,\,\,\,\,\,-\frac{1}{2}\beta (x)\beta^{T}{{(x)}}\frac{{{\partial }^{T}}J\left( \eta \right)}{\partial \eta}{{{\hat{W}}}_{av}}(t)-{{{\dot{\eta}}}_{d}}(t)\Big) ,
\end{align}
where ${{\alpha }_{c}}>0$ stands for the critic learning rate.
\begin{align}
		& {{{\dot{\hat{W}}}}_{a}}(t)=-\alpha \varphi (t)\Big({\Big(\delta {{g}^{T}}(x)\eta(t)+\frac{1}{2}g^{T}{{(x)}}\frac{{{\partial }^{T}}J( \eta)}{\partial \eta(t)}{{{\hat{W}}}_{a}}(t)\Big)^{T}} \cr
		& \,\,\,\,\,\,\,\,\,\,\,\,\,\,\,\,\,\times (g^{T}{{(x)}}\frac{{{\partial }^{T}}J(\eta)}{\partial \eta(t)}+(-\frac{1}{2}R(x)\frac{{{\partial }^{T}}J(\eta)}{\partial \eta(t)}) \cr
		& \,\,\,\,\,\,\,\,\,\,\,\,\,\,\,\,\,\times (2\delta \eta(t)+\frac{{{\partial }^{T}}J(\eta)}{\partial \eta}{{{\hat{W}}}_{c}}(t))\Big) ,
\end{align}
where ${{\alpha }}>0$ stands for the actor learning rate.
\begin{align}
	& {{{\dot{\hat{W}}}}_{av}}(t)=-{{\alpha }_{v}}\varphi (t)\Big(-{{\gamma }^{2}}(\delta \beta^{T}{{(x)}}\eta(t)-\frac{1}{2}\beta^{T}{{(x)}} \cr
	& \,\,\,\,\,\,\,\,\,\,\,\,\,\,\,\,\,\,\,\times \frac{{{\partial }^{T}}J(\eta)}{\partial \eta}W_{av}^{T}(t){{)}^{T}}(\beta^{T}{{(x)}}\frac{{{\partial }^{T}}J(\eta)}{\partial \eta}) \cr
	& \,\,\,\,\,\,\,\,\,\,\,\,\,\,\,\,\,\,\,+(-\frac{1}{2}\beta (x)\beta^{T}{{(x)}})\frac{{{\partial }^{T}}J(\eta)}{\partial \eta}(2\delta \eta(t) \cr
	& \,\,\,\,\,\,\,\,\,\,\,\,\,\,\,\,\,\,\,\,\,+\frac{{{\partial }^{T}}J(\eta)}{\partial \eta}{{{\hat{W}}}_{c}}(t))\Big),
\label{eq24}
\end{align}
where ${{\alpha }_{v}}>0$ stands for the actor learning rate for adversarial attacks.
\begin{remark}
  A reinforcement learning-based optimal control problem for a class of nonlinear systems under cyberattack is discussed in this paper. We have considered adversarial attacks in the proposed method in contrast to the works \cite{hr10,hr11}.
It can be seen from (\ref{eq15}) and (\ref{eq24}) that a matrix of neural network weights represents the impact of adversarial attacks, namely ${{W_{av}}^{*}}(t)$.
\end{remark}
It is possible to prove that the proposed optimized control problem is bounded and convergence is guaranteed. The weights of the neural networks must be shown to be convergent to achieve this goal. From the simulation result, it can be seen that the weights are effectively convergent. In order to demonstrate the stability of the proposed approach, we can use the Lyapunov function:
\begin{align*}
	& L(t)=\frac{1}{2}{{\eta}^{T}}(t)\eta(t)+\frac{1}{2}W_{a}^{T}(t)W_{a}(t) \cr
	& \,\,\,\,\,\,\,\,\,\,\,\,\,\,+\frac{1}{2}W_{c}^{T}(t){{W}_{c}}(t)+\frac{1}{2}W_{av}^{T}(t){{W}_{av}}(t).
\end{align*}
Then, for deriving the stability condition we should demonstrate that $\dot{L}(t)<0$. In order to accommodate space constraints within 6 pages and to improve the proposed method to consider some different scenarios, the authors are extending the proposed approach for journal publication.
\section{Simulation Results}
\label{sec3}
The proposed approach is tested using a robot manipulator to determine its performance.
\begin{figure}[H]
	\centering
	\includegraphics[width=0.7\linewidth,height=0.22\textheight]{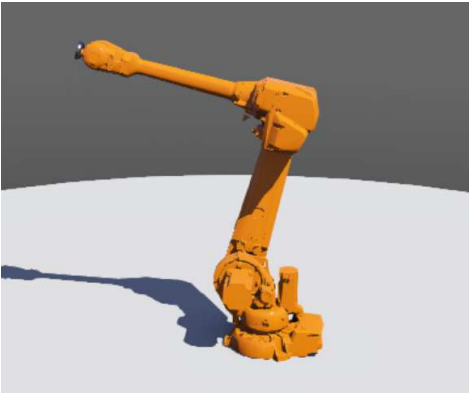}
	\caption{The manipulator in the Webots software environment..}
	\label{fig2}
\end{figure}
In this section, a robot manipulator as depicted in Fig. \ref{fig2},  its dynamics can be modeled as follows is presented to demonstrate the effectiveness of our proposed scheme:
\begin{align}
	\label{eq2771}
\bar{J}\ddot{q}(t)+B\dot{q}(t)+Mgl\sin \left( q(t) \right)=\tau(t).
\end{align}
In which $q(t)$ and $\dot{q}(t)$ represent angular and angular  velocity, respectively.   $\tau(t) $ stands for  the input torque. This is expressed as a function of the damping coefficient $B$, the total rotational inertia of the motor $\bar{J}$, the mass of the link $M$, the distance between the joint axis and the center of mass $l$, and the gravitational acceleration $g$. During the simulation, the parameter values are as follows: $\bar{J}=1$, $B=2,~Mgl=10$. Defining ${{x}_{1}}(t)=q(t),$ $~{{x}_{2}}(t)=\dot{q}(t)$, and $~u(t)=\tau(t) ,~$ the system (\ref{eq2771}) can be rewritten as follows:
\begin{align}
  & {{{\dot{x}}}_{1}}(t)={{x}_{2}}(t), \cr
 & {{{\dot{x}}}_{2}}(t)=-2{{x}_{2}}(t)-10\sin \left( {{x}_{1}}(t) \right)+u(t).
 \label{eq277}
\end{align}
In this case, assuming the system (\ref{eq277}) is subjected to adversarial attacks, it can be rewritten as follows according to equation (\ref{eq1}):
\begin{align*}
	\underbrace{\left[ \begin{matrix}
			{{{\dot{x}}}_{1}}(t)  \\
			{{{\dot{x}}}_{2}}(t)  \\
		\end{matrix} \right]}_{\dot{x}(t)}=\underbrace{\left[ \begin{matrix}
			{{x}_{2}}(t)  \\
			-2{{x}_{2}}(t)-10\sin ({{x}_{1}}(t))  \\
		\end{matrix} \right]}_{f(x)}+\underbrace{\left[ \begin{matrix}
			0  \\
			1  \\
		\end{matrix} \right]}_{g(x)}u(t)+\underbrace{\left[ \begin{matrix}
			1  \\
			1  \\
		\end{matrix} \right]}_{{\beta}(x)}v(t).
\end{align*}
The initial values are
$\text{sample time} ~~ T=0.1,
{{t}_{0}}=0,~{{t}_{f}}=20,~{{x}_{0}}={{\left[ 4~2 \right]}^{T}}$. In addition, the desired trajectory is set as ${{\dot{x}}_{d}}(t)={{\left[ \begin{matrix}
			{{{\dot{x}}}_{1d}}(t) & {{{\dot{x}}}_{2d}}(t)  \\
		\end{matrix} \right]}^{T}}=[\begin{matrix}
	{{x}_{2d}}(t) & 2\sin (0.7t)  \\
\end{matrix}]^T$
.
In what follows, we utilized a basis function vector neural network with 12 nodes for neural network approximation (\ref{eq12}).  Based on Gaussian function, the basis function vector is designed as
$S(x)={{[{{s}_{1}}(x),...,{{s}_{12}}(x)]}^{T}}$ with ${{s}_{^{i}}}(x)=\exp [-{{(x-{{\varsigma }_{i}})}^{T}}(x-{{\varsigma }_{i}})]$ . The neural network center is ${{\varsigma }_{i}}\in R,\,\,i=1,...,12$ equally spaced in the range $[-4,4]$ .
According to Figs. \ref{figs1}-\ref{figs5}, you can see the results of the simulation results of the proposed approach.
\begin{figure}[h!]
	\centering
	\includegraphics[width=0.98\linewidth,height=0.22\textheight]{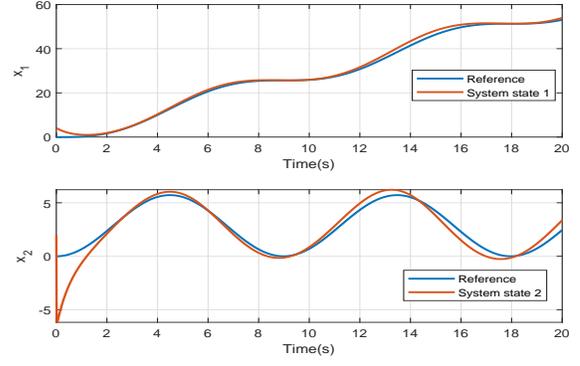}
	\caption{The trajectories of manipulator states.}
	\label{figs1}
\end{figure}\\
Based on the tracking performance displayed in Fig. \ref{figs1}, the proposed approach clearly allows the system to follow the desired trajectory with suitable performance.
\begin{figure}[h!]
	\centering
	\includegraphics[width=0.98\linewidth,height=0.22\textheight]{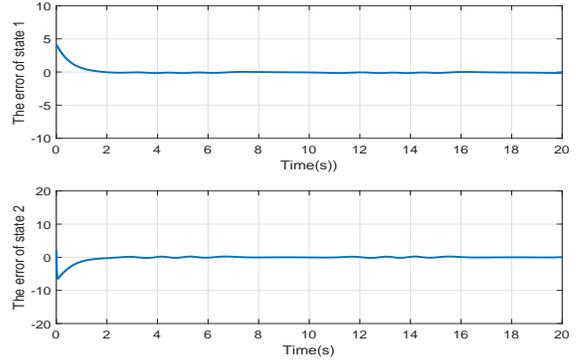}
	\caption{The tracking errors of manipulator states.  }
	\label{figs2}
\end{figure}
Fig \ref{figs2} shows the tracking errors of manipulator states between 0s and 20s.\\
Figs. \ref{figs3}-\ref{figs5} illustrate how neural network weights behave during simulation.  It takes the learner approximately two seconds to tune neural network weights to convergence.
\begin{figure}[h!]
	\centering
	\includegraphics[width=0.96\linewidth,height=0.21\textheight]{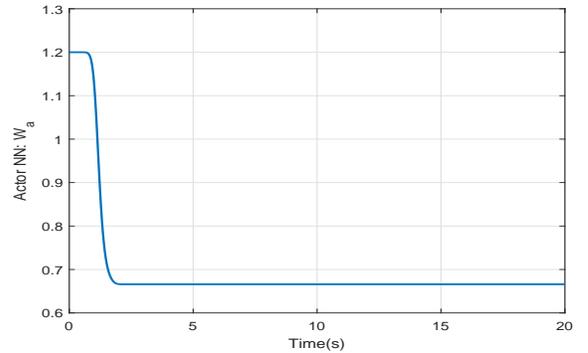}
	\caption{The convergence process of  ${W}_{a}(t)$. }
	\label{figs3}
\end{figure}
\begin{figure}[h!]
	\centering
	\includegraphics[width=0.96\linewidth,height=0.21\textheight]{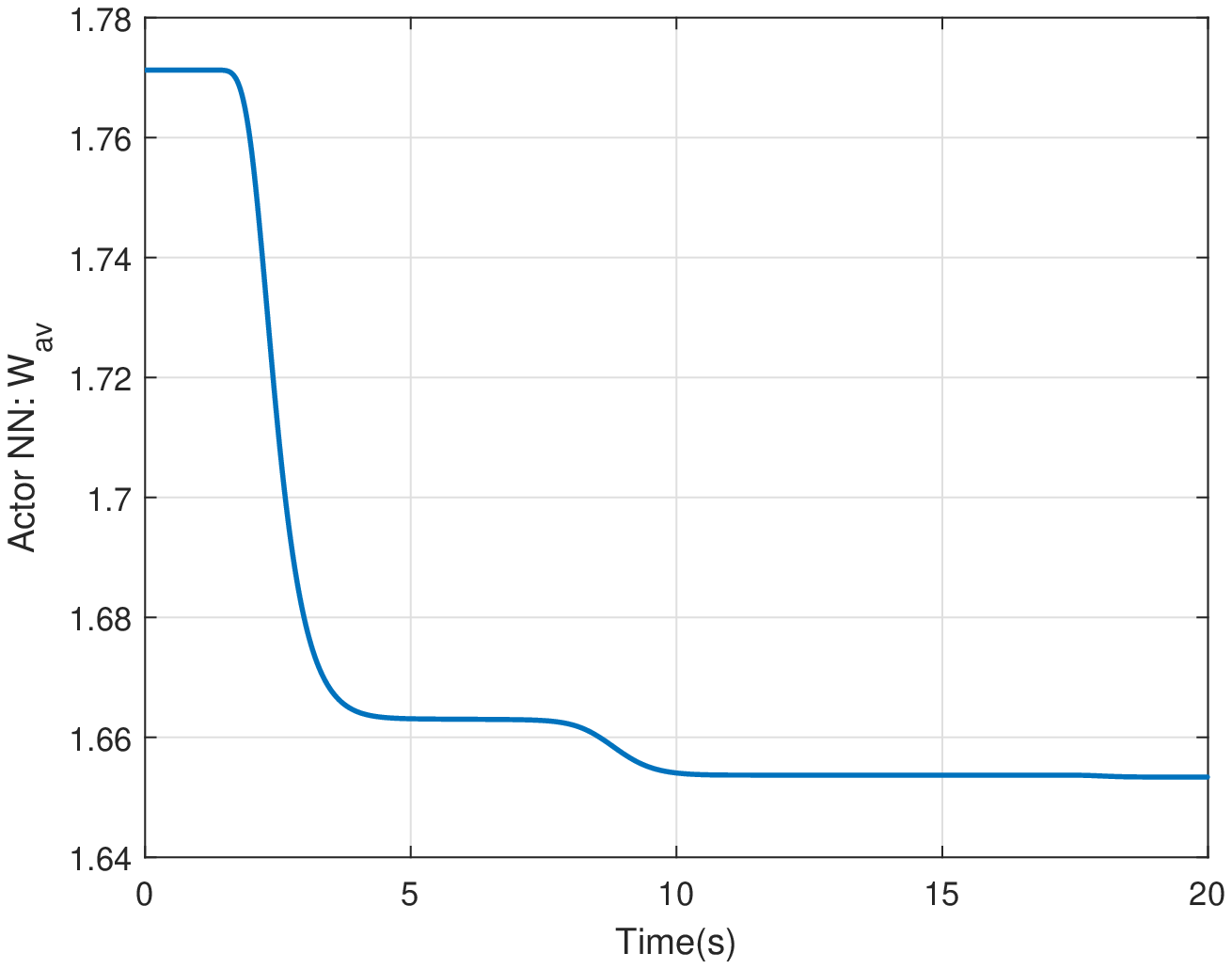}
	\caption{The convergence process of  ${W}_{av}(t)$.}
	\label{figs4}
\end{figure}
\begin{figure}[h!]
	\centering
	\includegraphics[width=0.96\linewidth,height=0.21\textheight]{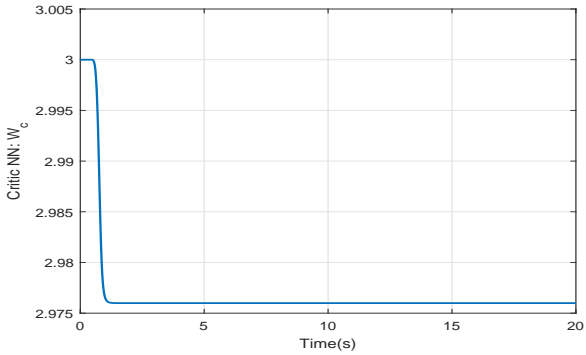}
	\caption{The convergence process of  ${W}_{c}(t)$.}
	\label{figs5}
\end{figure}
\section{Conclusion}
\label{sec4}
This study investigated a tracking control problem for a class of nonlinear systems by using neural networks based on reinforcement learning.  In our proposed approach to obtaining an optimal control input, adversarial attacks were considered.  Reinforcing learning was used to find the solution to the HJB equation in order to obtain optimal control inputs. Finally, simulation results verified the effectiveness of the proposed method for common robotic arms.  The proposed approach could be used to study multi-agent systems with quantization effects \cite{LPP1}; the problem of formation control for mobile robots; and detecting a cyber-attack or a fault in the system.

\bibliographystyle{IEEEtran}
\bibliography{Bib_Ref1}
\end{document}